\documentclass[pra,twocolumn,showpacs,amsmath,amssymb]{revtex4}
\usepackage{epsfig}
\usepackage{color}
\usepackage{graphicx}
\begin{document}

\title{Finite temperature phase diagram of spin-$1/2$ bosons in
two-dimensional optical lattice}
\author{L. de Forges de Parny$^1$, F. H\'ebert$^1$,
  V.G. Rousseau$^2$, and G.G. Batrouni$^{1,3,4}$}  
\affiliation{
$^1$INLN, Universit\'e de Nice-Sophia Antipolis, CNRS; 
1361 route des Lucioles, 06560 Valbonne, France,
}
\affiliation{
$^2$Department of Physics and Astronomy, Louisiana State University,
  B\^aton Rouge, Louisiana 70803, USA, 
}
\affiliation{$^3$Institut Universitaire de France}
\affiliation{
$^4$Centre for Quantum Technologies, National University of Singapore; 2
Science Drive 3 Singapore 117542.
}

\begin{abstract}
  We study a two-species bosonic Hubbard model on a two-dimensional
  square lattice by means of quantum Monte Carlo simulations and focus
  on finite temperature effects. We show in two different cases,
  ferro- and antiferromagnetic spin-spin interactions, that the phase
  diagram is composed of a superfluid phase and an unordered phase
  that can be separated into weakly compressible Mott insulators
  regions and compressible Bose liquid regions.  The superfluid-liquid
  transitions are of the Berezinsky-Kosterlitz-Thouless type whereas
  the insulator-liquid passages are crossovers.  We analyse the
  pseudo-spin correlations that are present in the different phases,
  focusing particularly on the existence of a polarization in this
  system.
\end{abstract}


\maketitle

\section{Introduction}

The study of strongly interacting quantum models by direct realization
of an experimental system reproducing the model properties, an idea
proposed by Feynman \cite{feynman82}, was realized in the past ten
years with the production of Bose-Einstein condensates (BEC) and their
use as ``quantum simulators" \cite{bloch08}. In particular, BEC in
conjunction with optical lattices are used to produce systems
reproducing the physics of well known quantum statistical discrete
models such as fermionic or bosonic Hubbard models.

Used to study simple models of bosons \cite{bloch02} or fermions
\cite{kett09} at low temperature, the flexibility offered by these
systems extends the range of interesting models to more exotic ones
which can be treated both experimentally and theoretically. Examples
include systems with long range interactions \cite{pfau05}, fermions
with imbalanced populations \cite{kett06}, mixtures of different kinds
of particles \cite{kett07} and spin-1 bosons with spin-independent and
spin-dependent interactions which allow interplay between
superfluidity and magnetism \cite{stamper, stamper2, ggb09}.
Furthermore, it is possible to study systems of bosons with two
effective internal degrees of freedom on an optical lattice, the
so-called ``spin-$1/2$ bosons". Such a system, with spin-dependent
interactions, could be produced by applying a periodic optical lattice
on a bosonic system with two triply degenerate internal energy levels.
The optical potential applied would localise the atoms at the nodes of
a regular network, but would also couple the internal states by
$\Lambda$ or V virtual processes, thus leaving only two internal low
energy degenerated states denoted $0$ and $\Lambda$ and realizing an
effective spin-$1/2$ model \cite{krutitsky04,larson}. The presence of
the spin-dependent interaction introduces a term in the Hamiltonian
which permits the conversion of two particles of one type into the
other type and renders numerical simulations more difficult. This
model is related, but not identical, to other models including p-band
superfluid models \cite{wu06a,wu06b,wu09} and the bosonic Kondo model
\cite{fossfeig}. Understanding the phase diagram and properties of the
simpler spin-$1/2$ bosonic system takes us a step toward understanding
the more elaborate, and more difficult to simulate, models.

The spin-$1/2$ model has been extensively studied with mean-field
theory (MFT) at zero or finite temperatures and, in previous work, we
explored its zero temperature behavior with quantum Monte Carlo (QMC)
simulations in one \cite{deforges10} and two \cite{deforges11}
dimensions for on-site repulsive interactions.  At zero temperature,
the phase diagrams obtained in one and two dimensions, with MFT or
QMC, are similar. Generally speaking, at zero temperature, the system
can adopt two different kinds of phases: insulating Mott phases that
appear for integer density, $\rho$, and for large enough repulsion
between particles and superfluid phases (SF) otherwise. The detailed
nature of these phases depends on the interactions between the
different kinds of particles.  In the case where the repulsion between
identical particles is smaller than between different particles ($U_2
> 0$ in our previous work \cite{deforges11} and in the following) the
superfluid is found to be polarized, that is, an imbalance develops in
the populations of the two kinds of particles and one of the species
becomes dominant. The $\rho=1$ Mott phase is also polarized whereas
the $\rho=2$ phase is not (we did not study higher densities with
QMC).  In the opposite case, $U_2 < 0$, all the phases are
unpolarized. Noteworthy is the presence of coherent exchange movements
\cite{kuklov03}, where two particles of different types exchange their
position in the $\rho=1$ Mott phase in both cases, as well as in the
$\rho=2$ Mott phase for $U_2< 0$.  Finally, in one dimension, all zero
temperature phase transitions were found to be continuous whereas in
two dimensions and when $U_2>0$ is small enough, the $\rho=2$
Mott-superfluid transition was predicted by MFT to be first order near
the tip of the Mott lobe and continuous otherwise, whereas for larger
$U_2$ the transition was predicted to be always continuous. This was
confirmed by QMC simulations \cite{deforges11}.  Related spin-1 models
were studied using MFT \cite{Pai08,kimura} or QMC in one dimension
\cite{ggb09} and a similar spin 1/2 bosons model was also recently
studied \cite{takayoshi10}.

In this paper, we will study the spin-$1/2$ model at finite
temperature in two dimensions and compare with MFT predictions. The
results for finite system sizes and temperatures are relevant to
experimental efforts to study this system. In Section II, we will
introduce the model and the MFT and QMC techniques used to study it.
Section III and IV will be devoted to the presentation of the results
obtained for the $U_2>0$ and $U_2<0$ cases, respectively. We will
summarize these results and give some final remarks in Section IV.

\section{Spin-1/2 Model\label{sec2}}

The model we will study is the same we previously studied in the low
temperature limit in one \cite{deforges10} and two dimensions
\cite{deforges11} and previously introduced in \cite{krutitsky04}. It
is an extended Hubbard model governed by the Hamiltonian
\begin{eqnarray}
&&\mathcal{H} = -t \sum_{\sigma, \langle {\bf r, r'} \rangle} 
\left(a^\dagger_{\sigma {\bf r}} a_{\sigma {\bf r'}}
+ a^\dagger_{\sigma {\bf r'}} a_{\sigma {\bf r}} \right) 
-\mu\sum_{\sigma, {\bf r}} \hat n_{\sigma {\bf r}} \label{h1}\\
&& +\,  \frac{U_0}{2} \sum_{\sigma, {\bf r}}\hat{n}_{\sigma {\bf r}}
(\hat n_{\sigma {\bf r}} -1 )
+ (U_0 + U_2) \sum_{\bf r} \hat{n}_{0 {\bf r}} \hat{n}_{\Lambda {\bf
    r}}   \label{h2}\\ 
&& + \frac{U_2}{2} \sum_{\bf r} \left( a_{0{\bf r}}^\dagger a_{0{\bf
    r}}^\dagger  a_{\Lambda {\bf r}}  
a_{\Lambda {\bf r}}
+  a_{\Lambda {\bf r}}^\dagger  a_{\Lambda {\bf r}}^\dagger  a_{0 {\bf
    r}} a_{0 {\bf r}} 
\right)\label{h3},
\end{eqnarray}
where operator $a_{\sigma{\bf r}}$ ($a^\dagger_{\sigma{\bf r}}$)
destroys (creates) a boson of type $\sigma = 0, \Lambda$ on site $\bf
r$ of a two-dimensional square lattice of size $L\times L$.  The
operators $\hat{n}_{\sigma {\bf r}}$ measures the number of particles
of type $\sigma$ on site $\bf r$.  The densities of particles of type
0 and $\Lambda$ are called $\rho_0$ and $\rho_\Lambda$ while the total
density is called $\rho=\rho_0 +\rho_\Lambda $.

The first term (\ref{h1}) of the Hamiltonian is the kinetic term that
lets particles hop from site $\bf r$ to its nearest neighbours $\bf
r'$.  The associated hopping energy $t=1$ sets the energy scale. A
chemical potential $\mu$ is added if one works in the grand canonical
ensemble. The second term (\ref{h2}) describes on-site repulsion
between identical particles with a strength $U_0$ or between different
particles with a strength $U_0 + U_2$. We will study both the positive
and negative $U_2$ cases but will keep only repulsive interactions,
that is $|U_2| < U_0$, and a fixed moderate value of $|U_2|/U_0 =0.1$.
The last term (\ref{h3}) provides a possibility to change the ``spins"
of the particles: When two identical particles are on the same site,
they can be transformed into two particles of the other type. It was
demonstrated in \cite{krutitsky04} that the matrix element associated
with this conversion is $U_2/2$.  We are using a different sign for
the term (\ref{h3}) compared to the articles where the model was
originally introduced \cite{krutitsky04} but we have shown in a
previous work \cite{deforges10} that this sign can indeed be chosen
freely due to a symmetry of the model.

\subsection{Mean Field Theory\label{mft}}

The only term that couples different sites in the Hamiltonian is the
hopping term (\ref{h1}).  Introducing the field $\psi_\sigma = \langle
a^\dagger_{\sigma {\bf r'}}\rangle = \langle a_{\sigma {\bf
    r'}}\rangle$, we replace the creation/destruction operators on
site ${\bf r'}$ by their mean values $\psi_\sigma$, following the
approach used in \cite{krutitsky04}. The Hamiltonian on site $\bf r$
is then decoupled from neighboring sites and can be easily
diagonalized numerically in a finite basis. The optimal value of the
fields $\psi_\sigma$ are then chosen by minimizing the grand canonical
potential $\mathcal{G} = -kT \ln {\cal Q}$ with respect to
$\psi_\sigma$ where ${\cal Q}$ is the grand canonical partition
function. The system is in a superfluid phase when $\psi_\sigma$ is
nonzero, with superfluid density $\rho_s = |\psi_0|^2 +
|\psi_\Lambda|^2$, and is otherwise in an unordered phase where two
cases can be distinguished: an almost incompressible case, {\it i.e.}
a Mott insulator, and a compressible case, {\it i.e.} a liquid.

In these two cases there is no broken symmetry and they cannot be
distinguished by symmetry considerations.  If there is a first order
transition between the MI and the liquid, characterized by
discontinuities in the density or other thermodynamic functions, the
MI and the liquid would be two distinct phases.  If, however, there is
no discontinuity in the evolution from the MI to the liquid then they
are only two limiting cases of the same unordered phase and there is
only a smooth crossover between the MI and the liquid regions of the
phase diagram. We shall see that, indeed, there is a crossover in the
system we are considering here.

One can distinguish between almost-incompressible and liquid regions
by calculating the local density variance which is a measure of the
local compressibility, ${\tilde \kappa} = \beta \left(\langle n_{\bf
  r}^2 \rangle - \langle n_{\bf r} \rangle^2\right)$ where $n_{\bf r}$
is the total number of particles on site $\bf r$. ${\tilde \kappa}$ is
close to zero in the Mott phase and much larger in the liquid phase.

While this MFT was shown to reproduce qualitatively the phase diagram
at zero temperature \cite{deforges11}, it is rather limited at finite
temperature. Indeed, whenever the $\psi_\sigma$ are zero the hopping
parameter $t$ no longer plays a r\^ole in the MFT. Then, while the MFT
can distinguish between SF and unordered phases, it does not correctly
distinguish Mott Insulator (MI) regions from normal Bose liquids ones, as
the crossover boundary between those regions will not depend on $t$
and will be the same as in the $t=0$ case.

\subsection{Quantum Monte Carlo simulations}

To simulate this system, we used the stochastic Green function
algorithm (SGF) \cite{SGF}, an exact Quantum Monte Carlo (QMC)
technique that allows canonical or grand canonical simulations of the
system at finite temperatures as well as measurements of many-particle
Green functions. In particular, this algorithm can simulate
efficiently the spin-flip term in the Hamiltonian. We studied sizes up
to $L=14$.  The density $\rho$ is conserved in canonical simulations,
but individual densities $\rho_0$ and $\rho_\Lambda$ fluctuate due to
the conversion term Eq. (\ref{h3}). The superfluid density is given by
fluctuations of the total winding number, $(W_0 + W_\Lambda)$, of the
world lines of the particles \cite{roy}
\begin{equation}
\rho_s = \frac{\langle (W_0 + W_\Lambda)^2\rangle}{4t\beta}.
\label{rhos}
\end{equation}
The superfluid density cannot be measured separately for $0$ and
$\Lambda$ particles due to the conversion term \cite{roscilde10}. It,
therefore depends on the total density not on the separate densities
of the two species.  We also calculate the one particle Green
functions
\begin{equation}
G_\sigma({\bf R}) =\frac{1}{2L^2}\sum_{\bf r} \langle a^\dagger_{\sigma {\bf
r+R}}a^{\phantom{\dagger}}_{\sigma {\bf r}} + a^\dagger_{\sigma {\bf
r}}a^{\phantom{\dagger}}_{\sigma {\bf
r+R}}\rangle,
\label{green1}
\end{equation}
with $\sigma=0, \Lambda$. $G_\sigma({\bf R})$ measures the phase
coherence of individual particles.

In a strongly correlated system it is useful to study correlated
movements of particles which can be done, for example, by studying
two-particle Green functions. We found \cite{deforges11} that
anticorrelated movements of particles govern the dynamics of particles
inside Mott lobes, as the particles of different types exchange their
positions. The two-particles anti-correlated Green function
\begin{eqnarray}
G_{\rm a}({\bf R}) &=&\frac{1}{2L^2} \sum_{\bf r} \left\langle
a^\dagger_{\Lambda {\bf r}}a^\dagger_{0{\bf r+R}}a^{\phantom{\dagger}}_{0{\bf
r}}a^{\phantom{\dagger}}_{\Lambda {\bf r+R}} + {\rm H.c.}
\right\rangle,
\label{greenf}
\end{eqnarray}
measures the phase coherence of such exchange movements as a function
of distance.  Due to its definition, $G_{\rm a}$ cannot be larger that
$\rho_0 \rho_\Lambda$ and is equal to $G_{\rm a}({\bf R}) = G_0({\bf
  R}) G_\Lambda({\bf R})$ if there is no correlation between the
movements of 0 and $\Lambda$ particles.  

In two dimensions, at low but finite temperatures, we expect, in some
cases, to observe a Berezinsky-Kosterlitz-Thouless (BKT) phase
transition and the different Green functions to adopt a power law
behavior at large distance ${\bf R}$, $G({\bf R}) \propto R^{-\eta}$,
characteristic of the appearance of a quasi long range order (QLRO) in
the phase, long range order (LRO) being achieved only at $T=0$ ($\eta$
varying between 1/4 and 0 as $T$ is lowered) \cite{lebellac}. In other
words, at finite $T$ we can expect a superfluid where the phase is
stiff but not ordered and not a BEC whith an ordered phase.  At high
temperature, the Green functions are of course expected to decay
exponentially.  On the finite size systems which we study ($L \le
14$), it is difficult to distinguish the QLRO from a true LRO, whereas
one can easily distinguish between the QLRO and exponentially
decreasing regimes. This difficulty of distinguishing QLRO from LRO is
also encountered in experiments where the sizes of systems that can be
studied are typically of the same order as in our QMC simulations
(hundreds of particles). The finite size results are, therefore,
directly relevant to experiments.

To elucidate the properties the model, we formulate it in terms of
spins using a Schwinger bosons approach \cite{auerbach}. Defining the
spin operators $S^z_{\bf r}=\left ({\hat n}_{0 {\bf r}}-{\hat
    n}_{\Lambda {\bf r}}\right) / 2$, $S^+_{\bf r} = a^\dagger_{0{\bf
    r}}a_{\Lambda{\bf r}}$, and $S^-_{\bf r} = a^\dagger_{\Lambda{\bf
    r}}a_{0{\bf r}}$ the Hamiltonian takes the form
\begin{eqnarray}
H&=& \hat{K}
+\,  \frac{U_0}{2} \sum_{\bf r}\hat{n}_{\bf r}
(\hat n_{\bf r} -1 ) + \frac{U_2}{4} \sum_{\bf r} \hat{n}^2_{\bf r}\label{HSpart1}\\ 
&+& U_2 \sum_{\bf r} \left( (S^x_{\bf r})^2 - [(S^y_{\bf
  r})^2+(S^z_{\bf r})^2]\right )\label{HSpart2}
\end{eqnarray}
where $\hat{K}$ is the hopping term Eq. (\ref{h1}).  The terms in
(\ref{HSpart1}) are invariant under spin rotations.  On the other
hand, term (\ref{HSpart2}) favors pseudo-spin correlations to develop
along the $x$ axis if $U_2 < 0$ or in the $yz$ plane for $U_2 > 0$. We
note that the total spin ${\bf S}^2_{\bf r} = s_{\bf r}(s_{\bf r}+1)$
on a given site is not fixed but depends on the total number of
particles on the site ($s_{\bf r} = n_{\bf r}/2$) and will then
fluctuate with this number.

An order along $z$ is measured through the densities or through
density-density correlations of 0 or $\Lambda$ particles, {\it i.e.}
it corresponds to the polarization of the system. An order along the
$x$ or $y$ axes is exposed through the behavior of $G_{\rm a}$ which,
in terms of spins, is equal to $G_{\rm a}({\bf R}) = \sum_{\bf r}
\langle S^x_{\bf r}S^x_{\bf r+R} + S^y_{\bf r}S^y_{\bf r+R}\rangle /
L^2$.  Our QMC algorithm allows the calculation of $G_{\rm a}({\bf
  R})$ but does not give access to correlations along the $x$ and $y$
axes independently.

We remark that the $y$ and $z$ axis have the same behavior, which
means that we expect a spin QLRO in the $yz$ plane to appear at low
enough temperature for $U_2 > 0$ in addition to the expected QLRO of
the global phase of the particles discussed earlier.  On our finite
size systems, this means that we should simultaneously observe a
polarization and a QLRO for $G_{\rm a}$. We will call such a
phenomenon a ``quasi-polarization" (QP).  For $U_2 > 0$ and finite
$T$, we expect a QLRO to develop along the $x$ axis and, consequently,
no polarization but still a plateau at long distances in the function
$G_{\rm a}$.

\section{$U_2 > 0 $ case}

At zero temperature, the results obtained with QMC and MFT were in
good qualitative agreement \cite{deforges11}. The phase diagram,
studied for densities up to two, exhibits three phases. The first two
phases are incompressible Mott phases obtained for integer densities
$\rho=1$ and $\rho=2$ for large enough interactions $U_0$. At zero
temperature, the entire $\rho=1$ Mott phase is polarized, that is the
system sustains a spontaneous symmetry breaking and the density of one
type of particles becomes dominant.  This polarization can be
understood in the framework of an effective spin-1/2 model
\cite{kuklov03} as the coupling in the $yz$ plane is stronger than
along the $x$ axis.  As expected, exchange movements of particles
coexist with polarization in this phase and $G_{\rm a}$ develop a long
range phase coherence. The $\rho=2$ Mott phase is unpolarized and show
no sign of exchange moves. In terms of spin, this corresponds to the
fact that, neglecting the kinetic term, the ground state for a given
site is uniquely determined as the state with $S^x_{\bf r} = 0$. The
third possible phase is a polarized superfluid (SF) and occurs at any
$U_0$ when the density is incommensurate and also at small $U_0$ for
commensurate values. It is not possible to discuss this phase in terms
of a simple effective spin degree of freedom as the number of
particles on a site is not fixed.

At zero temperature, the transition from the $\rho=1$ Mott phase to
the SF is continuous, whereas at the tip of the $\rho=2$ Mott lobe,
the transition to the SF is first order for small values of $U_2/U_0$
becoming second order for larger values.  This was predicted by the
MFT \cite{krutitsky04} and confirmed by QMC simulations
\cite{deforges11}.

\subsection{Phase diagram at $T \ne 0$}

\begin{figure}
\includegraphics[width=8.5cm]{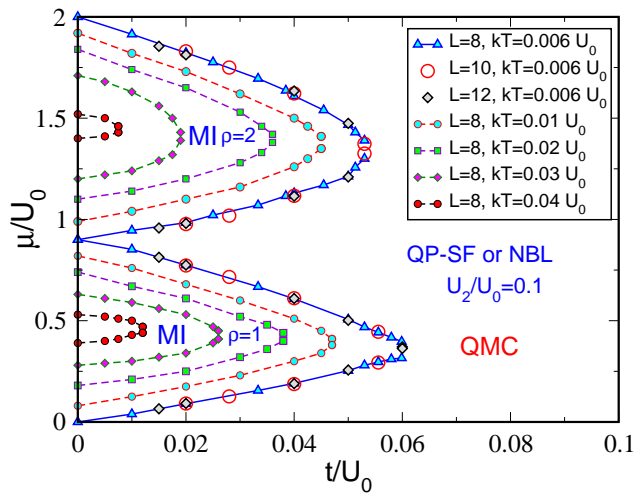}
\caption{(Color online) The limits of the $\rho=1$ and $\rho=2$ Mott
  regions (MI) for different values of the temperature
  $kT$ and for different sizes $L$. As the temperature increases, the
  Mott lobes shrink and totally disappear for $kT > 0.04 \,
  U_0$. Outside the Mott, the system crosses over to a normal Bose
  liquid (NBL) and, eventually, 
  transitions to a quasi-polarized superfluid (QP-SF) phase (see
  below). In the limit of zero  
  temperature (here represented by the low temperature $kT/U_0 =
  0.006$), there is a direct phase transition between a Mott phase and
  a superfluid.
  \label{mottlimits}}
\end{figure}

\begin{figure}[b]
\includegraphics[width=8.5cm]{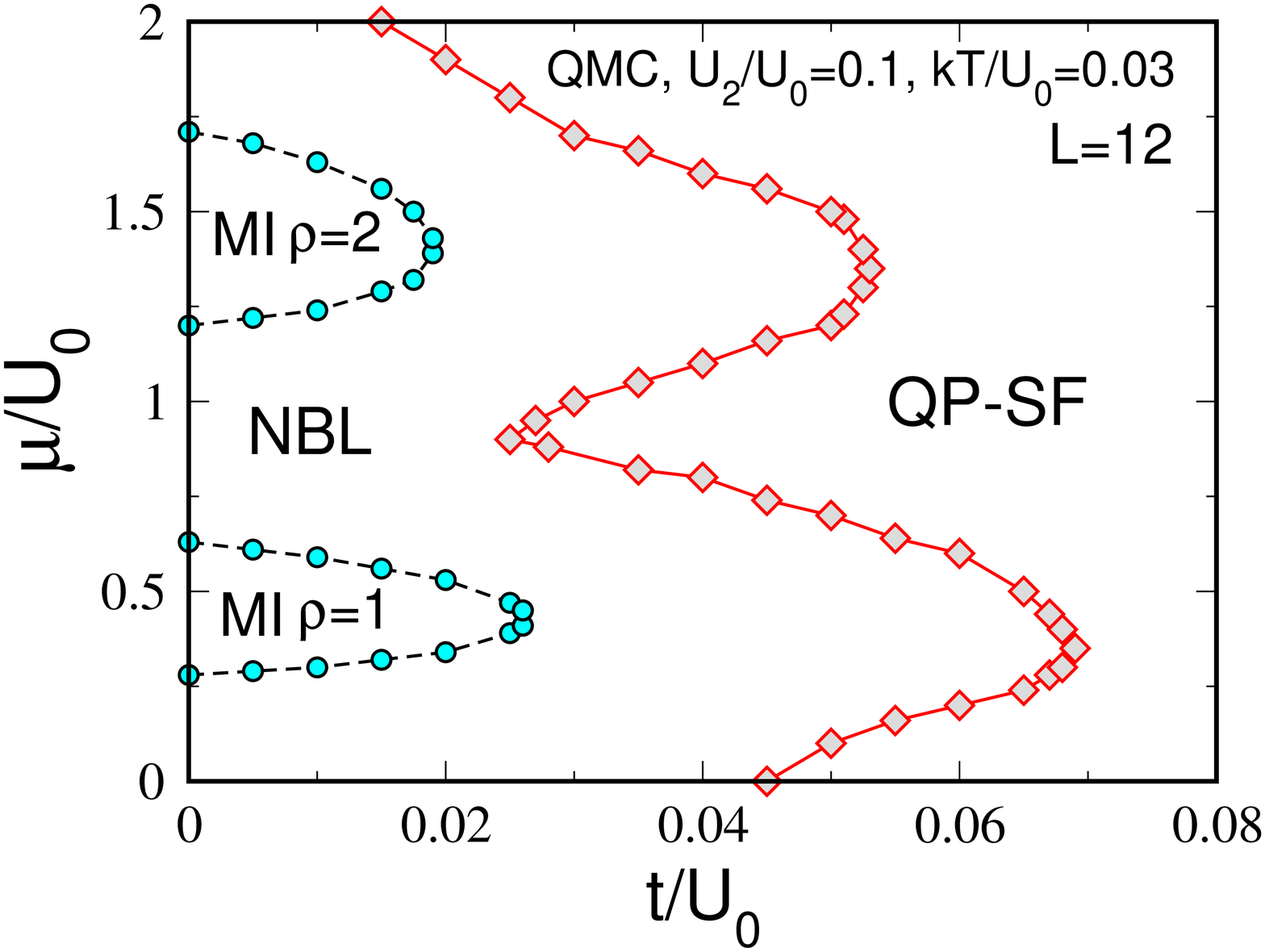}
\includegraphics[width=8.5cm]{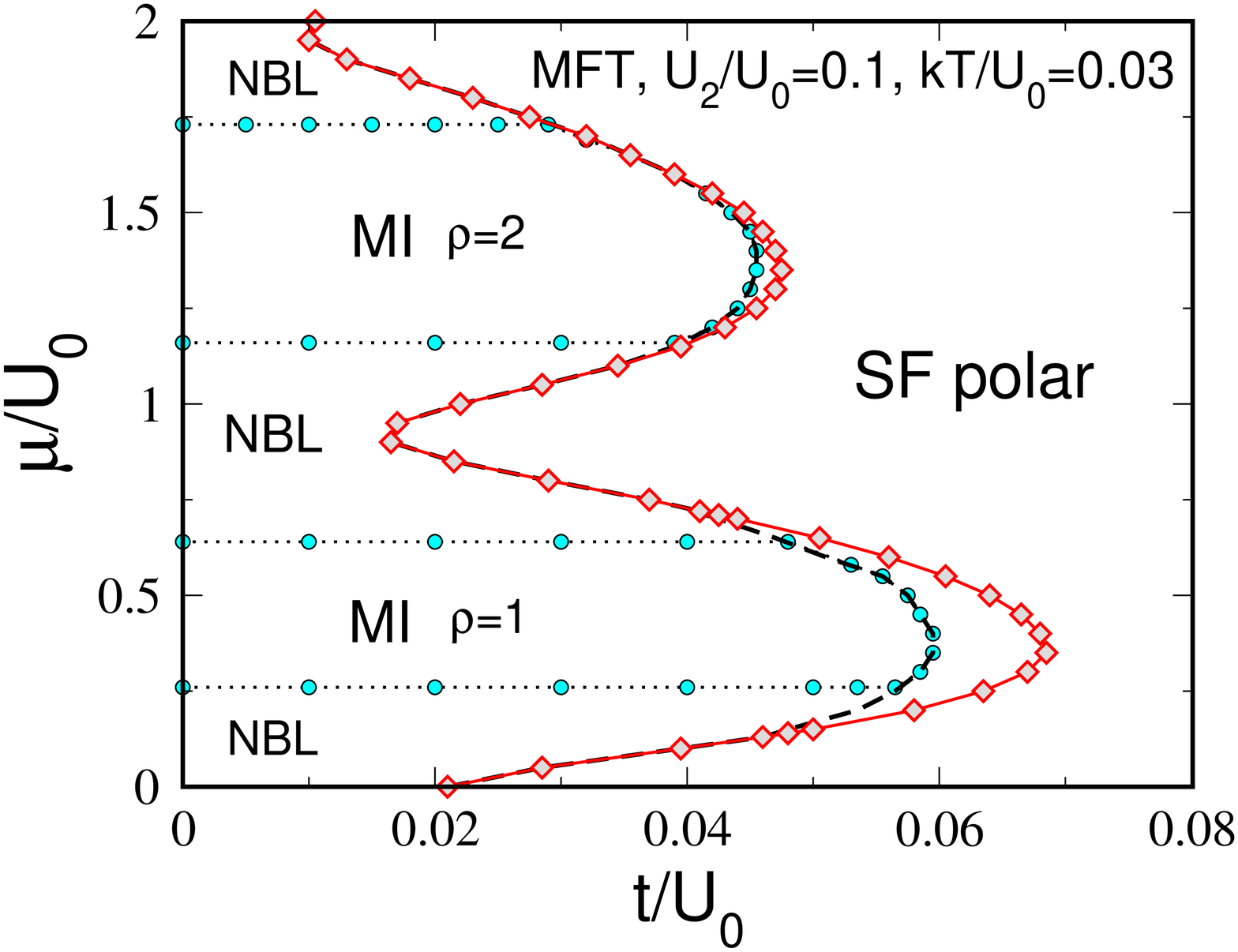}
\caption{(Color online) Phase diagram of the system at $kT=0.03\, U_0$
  for $U_2/U_0=0.1$.  The QMC simulations (top) show that a liquid (NBL) region
  appears between the quasi polarized superfluid (QP-SF) phase and the Mott
  region. The MFT (bottom) is not able to reproduce
  correctly this result as it does not take into account the kinetic
  term in the unordered (Mott or NBL) phase. It predicts
  first order phase transition between the SF and the other phase at
  the tip of the $\rho=1$ and $\rho=2$ lobes: the region of
  coexistence of the two phases is limited by the dashed line and the
  redline with squares. 
  \label{phasediagram}}
\end{figure}

To map the phase diagram at finite temperature with QMC, we determine
the limit of the Mott Insulator regions by measuring the density as a
function of $\mu$ and determining the boundaries of the plateaux
indicating the almost incompressible regions.  As explained in
Sec. IIA, although the compressibility of these regions is very small,
they are not strictly incompressible due to thermal fluctuations.  The
evolution from MI to NBL does not show any singularity and is then
simply a crossover between two different limiting behaviors,
incompressible and compressible, of the same unordered phase.
Strictly speaking, a truly incompressible Mott phase exists only at
$T=0$ but, following convention, we will continue to refer to this
finite $T$ region as a MI.  To define the crossover boundary between
the MI and the NBL shown in phase diagrams, we use the following
criterion: when the density deviates by 1\% from the total integer
density we consider that the system is no longer in the MI region but
in the NBL region.

Figure \ref{mottlimits} shows the limits of the MI
regions for different temperatures. As expected, the  MI
regions progressively disappear as the temperature is increased.  The
limits of the superfluid phase are located by direct measurement of
$\rho_s$. When the system is compressible and has zero superfluid
density, there is a normal Bose liquid.
Fig.~\ref{phasediagram} (top) shows the QMC phase diagram of
the system at a constant finite temperature. As the SF and
MI regions are destroyed due to thermal fluctuations,
an intermediate NBL region appears.  The
MFT used with success at $T=0$, where it reproduces qualitatively the
phase diagram, is unable to do so at finite temperature as explained
in Sec. \ref{mft}. Figure~\ref{phasediagram} (bottom) shows the MFT
phase diagram where the afore mentioned problem clearly appears: the
boundaries between the MI and NBL do not depend on the
value of $t/U_0$ and the MFT is unable to give correct predictions
regarding this crossover. However, the boundaries of the SF are
reasonably well reproduced. A surprising result is that the
transition between the unordered phase and the SF appears to be
first order at the tip of the $\rho=1$ and $\rho=2$ lobes in this MFT
approach. At zero temperature, only $\rho=2$ showed a first order
transition. As will be shown below, for finite temperatures, this is
in total contradiction with the results obtained by QMC simulations
that show continuous phase transitions for $\rho=1$ and $\rho=2$. So
this MFT provides incorrect description of the phase transition and of
the position of the different regions at finite temperature.

\begin{figure}
\includegraphics[width=8.5cm,clip]{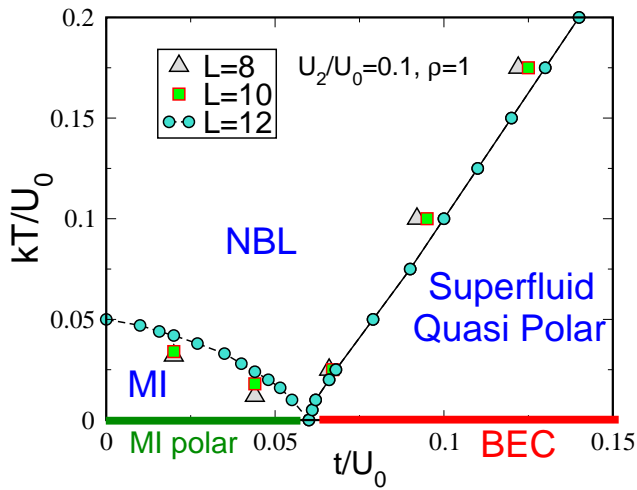}
\caption{(Color online) The QMC phase diagram for $\rho=1$ and for
  different sizes showing the quantum phase transition point between
  the Mott insulating  phase (MI) and the superfluid at
  $kT=0$. The finite $T$ superfluid phase
  does not have BEC in this two-dimensional system. The
  system is quasi-polarized throughout the superfluid phase, but it is not
  in the NBL and MI. The Mott phase is polarized at $T=0$ and should be
quasi-polarized at extremely small $T$. The
  transition between the SF and the NBL is continuous and there
  is only a crossover separating NBL from MI.
  \label{afrho1}}
\end{figure}

\begin{figure}
\includegraphics[width=8.5cm,clip]{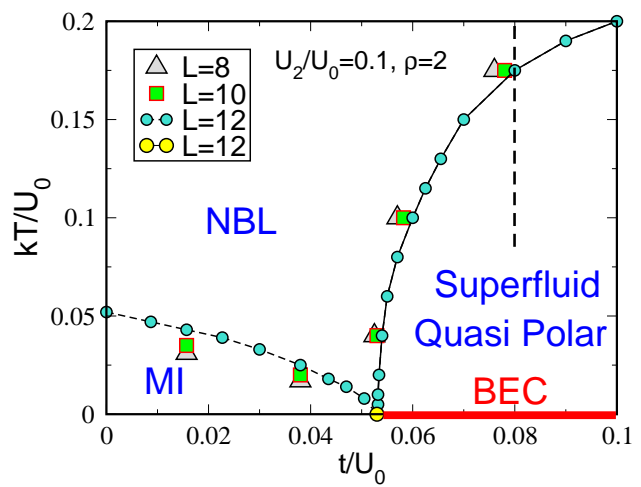}
\caption{(Color online) The QMC phase diagram for $\rho=2$ and for
  different sizes showing the quantum phase transition point between
  the MI and the SF occurring only at $kT=0$. The superfluid phase is
  always quasi-polarized whereas the NBL and MI are not. The SF-NBL
  transition is continuous, whereas at $T=0$ the MI-SF
  quantum phase transition was found to be discontinuous for $U_2/U_0$
  small enough \cite{deforges11}. The dash line corresponds to the 
case studied in Fig.~\ref{histo}.}
\label{afrho2}
\end{figure}

For fixed integer density at zero temperature, we have a quantum phase
transition (QPT) between the MI and SF phases.  
As expected \cite{sachdev}, when the temperature is increased from zero,
an intermediate compressible unordered region appears between the
superfluid phase and the Mott region, namely the normal Bose liquid region (NBL). 
This is observed for $\rho=1$ (Fig.~\ref{afrho1}) and $\rho=2$
(Fig.~\ref{afrho2}).

\begin{figure}
\includegraphics[width=8.5cm]{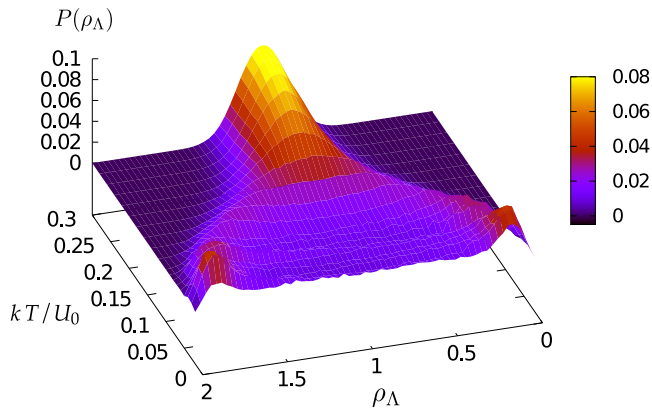}
\caption{(Color online) Probability $P(\rho_\Lambda)$ as a function of
  temperature for $L=8$, $U_2/U_0 = 0.1$, $\rho=2$ and $t/U_0=
  0.08$. When $kT/U_0 < 0.175$, the system is in the quasi-polarized
  SF phase. For $kT/U_0 > 0.175$, the system is in the unordered
  unpolarized phase.  The temperature at which the quasi-polarization
  disappears is approximately the same as the temperature where
  $\rho_s$ becomes zero (see the vertical dashed line in
  Fig.~\ref{afrho2}). Only the SF phase is
  quasi-polarized.\label{histo}}
\end{figure}

As for the possible polarization of the different phases, we observe
that the SF phase appears to be polarized at finite $T$ as it is at
$T=0$: the histogram of the density of one of the species shows two
peaks at low and high densities (see Fig. \ref{histo}).  However, due
to the continuous symmetry in the $yz$ plane of the pseudo-spin part
of the Hamiltonian Eq. (\ref{HSpart2}), no LRO exists at finite
$T$. Therefore, this apparent polarization is due to the finite size
of the system, and is, in fact, a quasi-polarization.  An intuitive
way to understand this quasi-polarization is as follows: In the
superfluid phase, the pseudo-spins in the $yz$ plane are stiff and
appear to be mostly aligned in the same direction on a finite lattice,
such as the case here. But since the symmetry is not broken, this
``magnetization'' direction in the $yz$ plane will drift and point in
all directions. As the direction changes, so does the projection of
this pseudo-spin on the $z$-axis. In other words, the polarization
drifts too, and changes with time, giving the double peak structure to
the polarization histogram, Fig. \ref{histo}. We note that
experimental systems are typically of the same sizes as the ones we
study here and, therefore, this polarization drift will be present in
these experiments too: The particle content of the system will change
as a function of time. This quasi-polarization disappears as $T$
increases when the system undergoes a thermal BKT phase transition
into the NBL. In other words, the entire SF phase is quasi-polarized
but the NBL is not (see Fig.  \ref{histo})

\begin{figure}
\includegraphics[width=8.5cm]{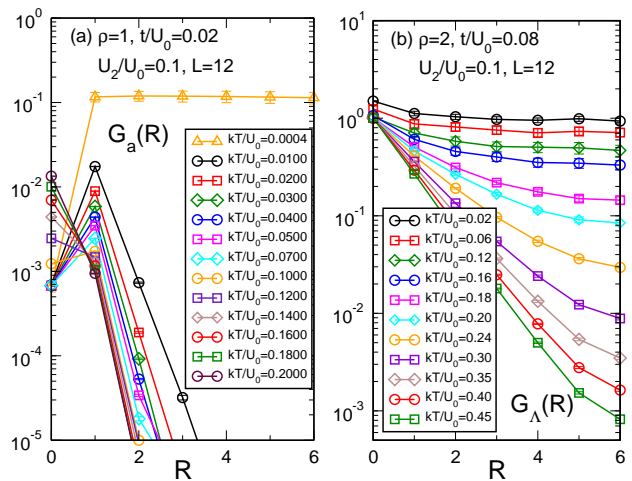}
\caption{(Color online) Evolution of the Green functions $G_{\rm a}$,
  Eq. (\ref{greenf}), in the $\rho=1$ MI and $G_\Lambda$,
  Eq. (\ref{green1}), in the SF as functions of distance $\bf R$ for
  different temperatures, with $L=12$ and $U_2/U_0=0.1$. (a) In the
  $\rho=1$ MI, $G_{\rm a}$ decays exponentially at finite $T$, unlike
  its $T=0$ behavior where it reaches a constant value
  \cite{deforges11}. A regime of QLRO
is not observed and would occur only at very small $T$. The
exponential decay persists in the NBL at 
  higher $T$.  (b) $G_\Lambda$ at $\rho=2$ for $T$ values taking the
  system from the SF to the NBL. Same case as in Fig. \ref{histo}. In
  the SF, $G_\Lambda$ decays slowly 
  as a power law, but decays exponentially in the NBL. Similar behavior
  is found for $G_0$ and $G_{\rm a}$. 
\label{green}
}
\end{figure}

A histogram of the polarization in the $\rho=1$ MI region shows that
as soon as the temperature is increased from zero, the polarization
disappears and the populations become balanced.  According to the
effective pseudo-spin model \cite{kuklov03, deforges11}, it is
possible to observe quasi-polarization as in the SF phase.  In this
case, the coupling generating these spin correlations and the
polarization of the MI is of order $t^2/U_0$. However, even for
temperatures as low as $kT/U_0=0.01$, the system is already in the
regime where correlations decay exponentially and we do not observe
any QLRO at finite $T$ in the Mott region. This is confirmed by the
behavior of the Green functions that is shown in Fig. \ref{green}. In
Fig. \ref{green}(a) we show the anticorrelated Green function $G_{\rm
  a}$, Eq. (\ref{greenf}), in the $\rho=1$ MI region as $T$ is
increased. We see that as soon as $T$ becomes finite, $G_{\rm a}$
decays exponentially with distance, contrary to its constant value at
long distance observed at $T=0$ \cite{deforges11}. This exponential
decay of course persists in the NBL region.

In Fig. \ref{green}(b) we show
$G_{\Lambda}(R)$, Eq. (\ref{green1}), for various $T$ values at
$\rho=2$ and $t/U_0=0.08$ which, at $T=0$, puts the system in the SF
phase. We see that for low $T$, $G_\Lambda$ decays slowly. This decay
is expected to be a power law but the system size is too small to show
that unambiguously. As $T$ is increased, the system transitions into
the NBL where $G_\Lambda$ exhibits exponential
decay clearly distinguishing the NBL and SF phases. A similar behavior
is found for $G_0$ and $G_{\rm a}$, the latter being expected since 
it accompanies the presence of quasi-polarization.

\subsection{Nature of the transitions}

\begin{figure}
\includegraphics[width=8.5cm]{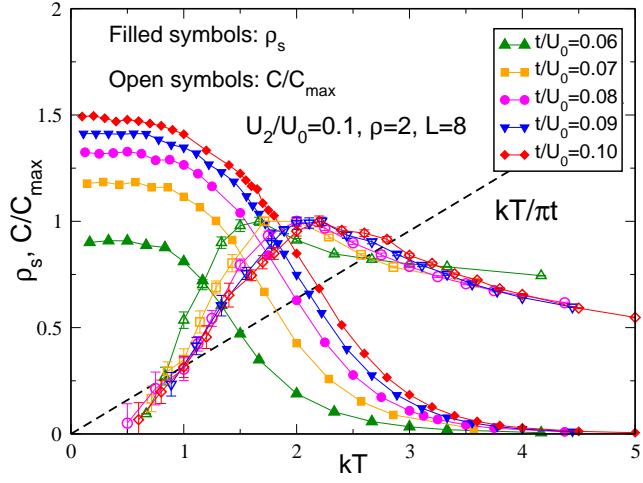}
\caption{(Color online) Superfluid density $\rho_s$ and specific heat
  $C$ as functions of temperature for different values of $t/U_0$. The
  universal jump condition states that $\rho_s(T_c)= kT_c/\pi t$. The
  transition temperatures are calculated independently by determining
  the maximum of $C(T)$.
\label{univjump}}
\end{figure}

At finite temperature, $kT > 0.005\, U_0$, the transitions between the
SF and the NBL are continuous for $\rho=1$ as well as for $\rho=2$
(and of course for all other densities). Since this transition takes
place at constant density and is therefore a phase-only transition, it
is expected to be in the Berezinsky-Kosterlitz-Thouless universality
class (BKT) for our two-dimensional system. Actually, below the
critical temperature, we have two different quasi long range orders
that occur: the global $U(1)$ phase QLRO associated with the
superfluid behavior and the pseudo-spin one, associated with the
quasi-ordering of the spins in the $yz$ plane, {\it i.e.} the
so-called quasi polarization. As explained previously, these two QLRO
appear simultaneously. Since the transition is BKT, we first
determined the transition temperature using the universal jump of the
superfluid density \cite{nelson77}, where, at the transition
temperature, $T_c$, we have $\rho_s(T_c)=kT_c/\pi t$. To observe this,
we calculated $\rho_s$ as a function of temperature and determined
$T_c(\rho_s)$ graphically as the intersection of $\rho_s(T)$ with
$kT/\pi t$ (see Fig. \ref{univjump}). We also calculated the specific
heat $C$ and determined the transition temperatures $T_c(C_{\rm max})$
as the temperature where $C$ reaches its maximum
(Fig. \ref{univjump}). Our QMC simulations were done for $L=8$ because
$C$ is extremely difficult to calculate at low temperature for $U_2
>0$ for larger sizes.

\begin{table}
\begin{tabular}{ccc}
\hline
$t/U_0$ & $T_c \ (\rho_s)$ & $T_c \ (C_{\rm max}) $ \\
\hline
0.06 & $1.49  \pm 0.10 $ & $1.67 \pm 0.15$ \\
0.07 & $1.82  \pm 0.05 $ & $1.85  \pm 0.15$ \\
0.08 & $1.99  \pm 0.05 $ & $2.00  \pm 0.15$ \\
0.09 & $2.10  \pm 0.05 $ & $2.10  \pm 0.20$\\
0.10 & $2.19  \pm 0.05 $ & $2.20  \pm 0.20$\\
\hline
\end{tabular}
\caption{Values of the transition temperatures for different values of
  $t/U_0$ for an $8\times 8$ system.  $T_c(\rho_s)$ is obtained
  from the universal jump while $T_c(C_{\rm max})$ from the maximum of
  $C(T)$. The values are in agreement and show that the universal
  prediction is verified. The data are taken from
  Fig.~\ref{univjump}. \label{tab1}}
\end{table}

\begin{figure}
\includegraphics[width=8.5cm]{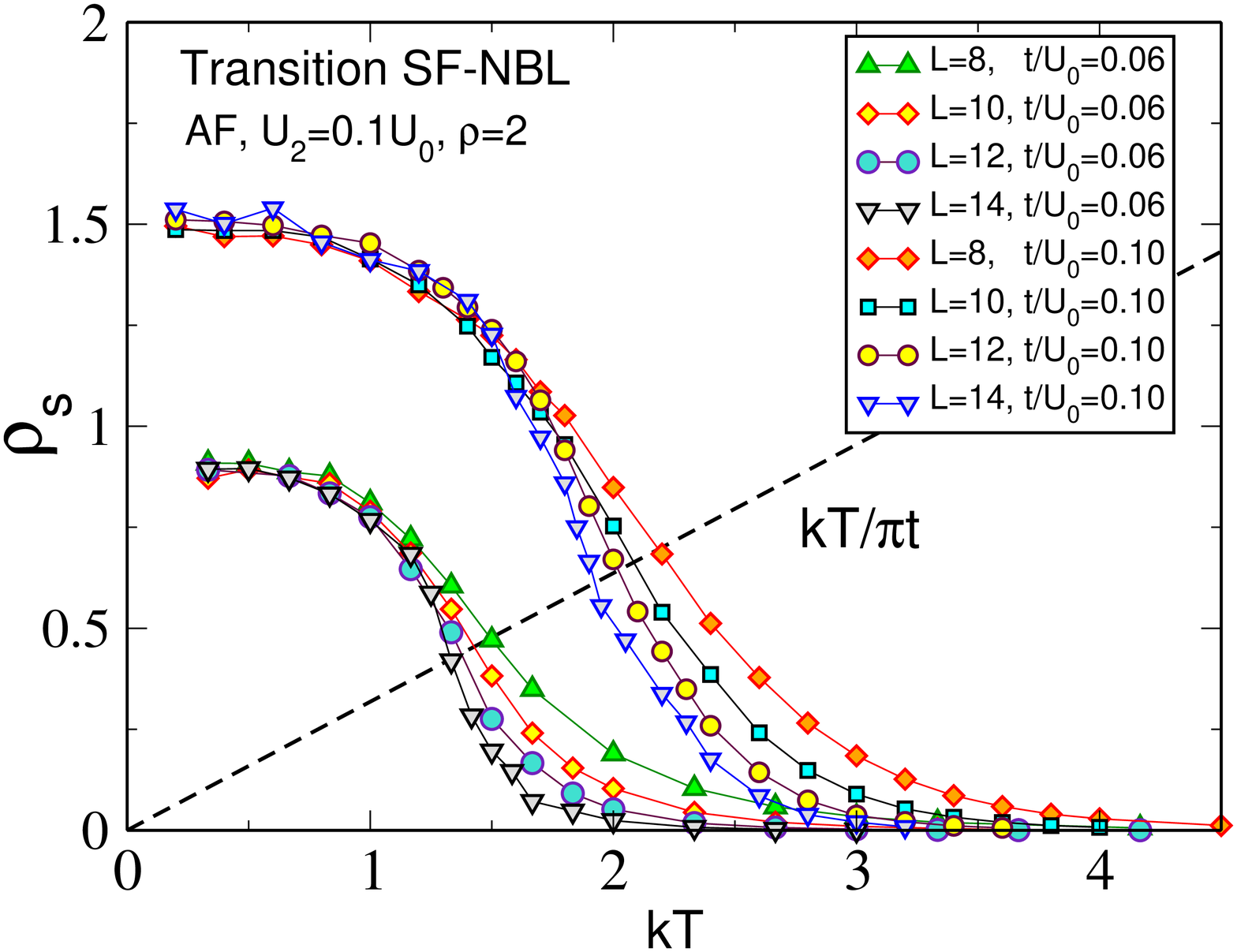}
\caption{(Color online) Superfluid density $\rho_s$ as a function of
  $T$ for $t/U_0=0.06$ and $0.10$ and for different sizes $L$. The
  transition becomes sharper as $L$ is varied from $L=8$ to
  $L=14$.\label{univ2}}
\end{figure}

Table \ref{tab1} compares the values of $T_c$ obtained from the
universal jump and from the maximum of $C$. The values are in
agreement confirming the universal jump hypothesis and the BKT nature
of the transition and determining the transition temperature for the
studied size. We did simulations for sizes up to $L=14$ to
examine the effect of finite size for this transition (see
Fig. \ref{univ2}). As expected, the transition gets sharper with
increasing size. However, it is too difficult to obtain results with
small enough error bars for $C$ on these large sizes.

\begin{figure}
\includegraphics[width=8.5cm]{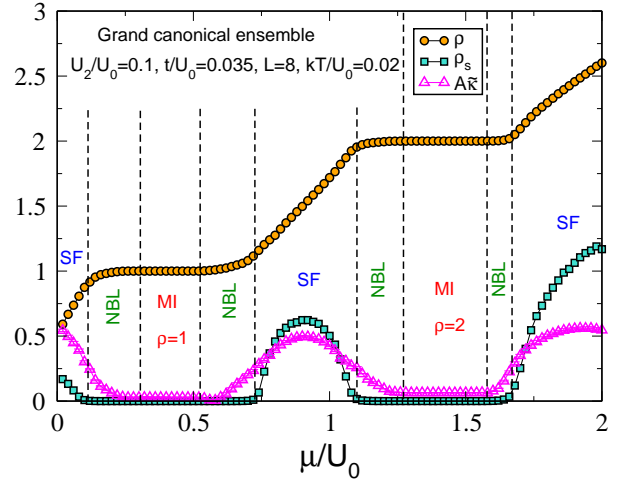}
\caption{(Color online) $\rho$, $\rho_s$ and ${\tilde \kappa}$ as
functions of $\mu$. As one goes from the MI to the NBL, the density
varies continuously indicating a crossover between the two
regions. The two regions are distinguished by the integer/non-integer
value of the density and by the larger value of the local
compressibility ${\tilde\kappa}$ in the liquid. The superfluid phase
has nonzero $\rho_s$ and a much larger local compressibility than both
the MI and NBL.  ${\tilde \kappa}$ has been multiplied by an arbitrary
factor $A$ for better visibility.
\label{rhovsmu}}
\end{figure}

As mentioned earlier, the evolution from the  MI region to the NBL is a
continuous crossover.  A plot of the density as a function of $\mu$ shows no
sign of a first order phase transition in the form of a jump in the
density as one approaches the Mott plateaux (see Fig. \ref{rhovsmu}).
There is no phase transition between MI and NBL since, in addition to
the absence of a first order transition, no symmetries are
broken. Then, at finite temperature, there is only a crossover between
the MI and the NBL and not the phase transition predicted by MFT.

\begin{figure}
\includegraphics[width=8.5cm]{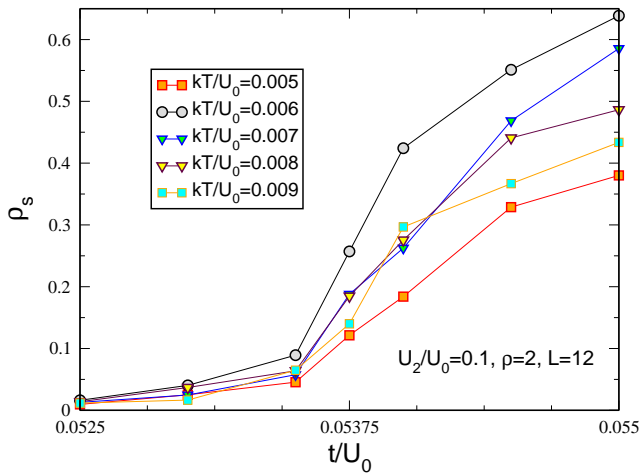}
\caption{(Color online) Transition from the normal Bose liquid to the superfluid phase at low
  temperatures. The transition always appears continuous for low
  temperatures. The discontinuity is only found in the zero
  temperature regime ($kT < 10^{-3}U_0$). \label{discontinu}}
\end{figure}

At zero temperature, we have shown \cite{deforges11} that the Mott-SF
transition is always second order for $\rho=1$ but is first order near
the tip of the $\rho=2$ Mott lobe when $U_2/U_0$ is small enough, for
example $U_2/U_0=0.1$. Hence, while it is easy to imagine that the
continuous NBL-SF transitions observed at moderate temperatures
persists at low temperature for $\rho=1$, the $\rho=2$ case where the
behavior is different at zero and finite temperatures requires a
separate study for the low temperature regime.  Probing temperatures
as low as $kT = 0.005 U_0$ (see Fig.  \ref{discontinu}), the
transition still appears continuous, which is to be compared with the
temperature at which the MI region  is destroyed $kT \approx 0.05 U_0$
(see Figs. \ref{afrho1} and \ref{afrho2}). Although it cannot be
excluded that the transition is discontinuous in a small range of
temperatures, this range would be extremely narrow. In order to
observe the first order QPT in our previous work \cite{deforges11} we
used temperatures that were of order $kT/U_0 \simeq 10^{-3}$. The
transition then appears discontinuous only at extremely low
temperatures.

\section{$U_2 < 0$ case}

It was shown for $U_2 < 0$ \cite{deforges11} that all the phases are
unpolarized and that transitions between the different phases are all
continuous in the $T=0$ limit. The coherent anticorrelated movements
are present in the $\rho=1$ and in the $\rho=2$ Mott phases.

\begin{figure}
\includegraphics[width=8.5cm]{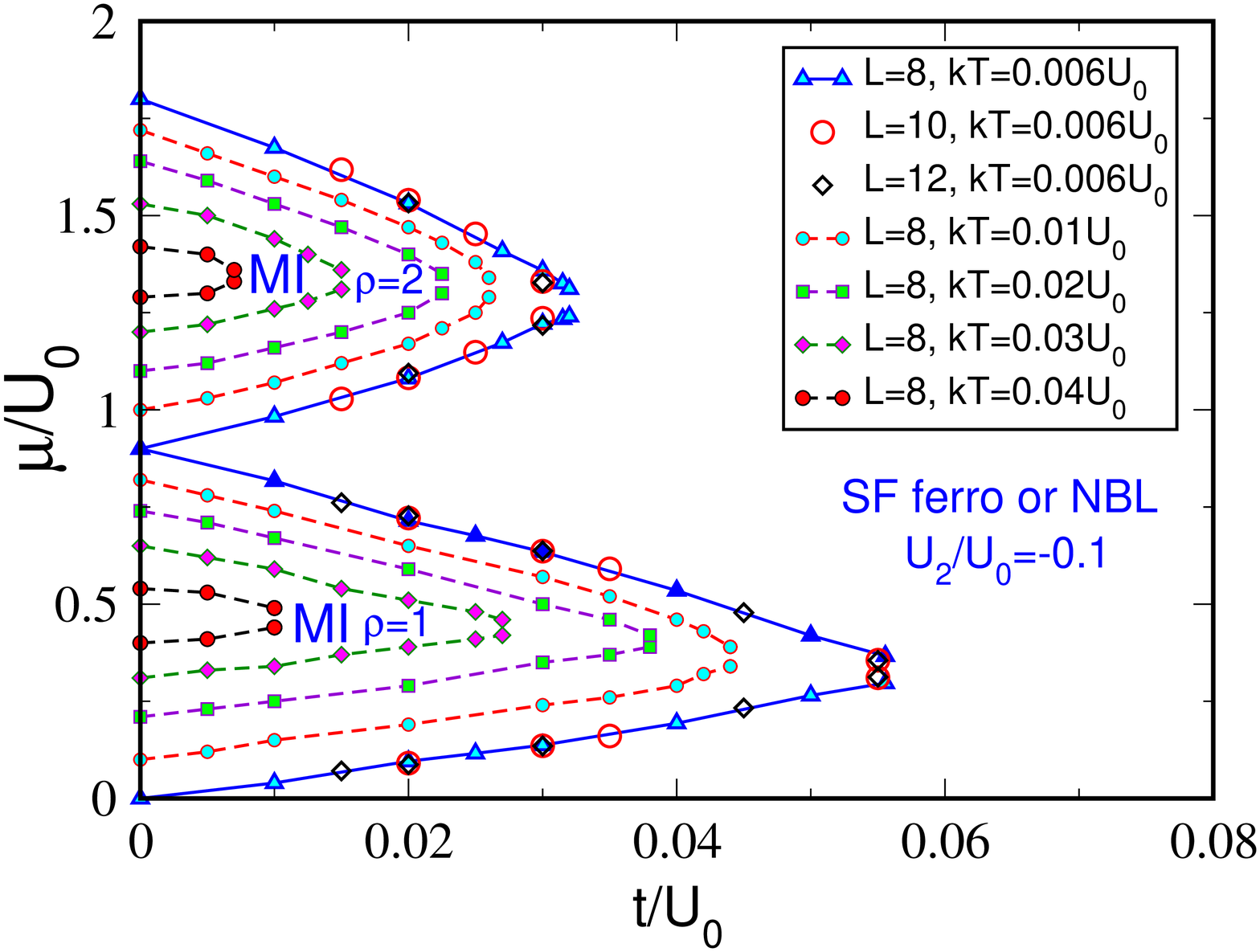}
\caption{(Color online) The limit of the $\rho=1$ and $\rho=2$ Mott
  regions (MI) for different values of the temperature $kT$ and for
  different sizes $L$ in the $U_2 < 0$ case. As in the positive $U_2$
  case, the Mott phases disappear for $kT > 0.04 U_0$. In the $T=0$
  limit, there is a direct transition between the MI and
  SF.\label{mottlimit2}}
\end{figure}

\begin{figure}
\includegraphics[width=8.5cm]{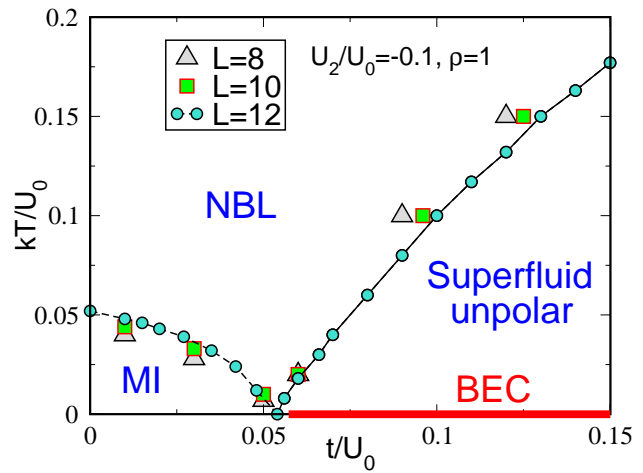}
\caption{(Color online) The phase diagram for $\rho=1$ in the $U_2 <
  0$ case. All phases are unpolarized and the SF-NBL
  transition at finite temperature as well as the MI-SF transition at
  zero temperature are continuous. A similar phase diagram is
  observed for $\rho=2$.\label{rho12_U2neg}}
\end{figure}

Proceeding as in the $U_2 > 0$ case, we determine the boundaries of
the Mott regions at finite $T$, Fig.~\ref{mottlimit2}, by calculating
the density and the boundary of the superfluid region by measuring
$\rho_s$. In this case, we chose not to present results from MFT since
it shows the same limitations as in the $U_2 > 0$ case. We obtain the
phase diagram for $\rho=1$ (shown in Fig.  \ref{rho12_U2neg}) and a
similar one for $\rho=2$ (not shown here).  Similarly, we used
histograms of the density similar to Fig. \ref{histo} to confirm that
all these phases remain unpolarized at finite temperature as they are
at zero temperature. The absence of polarization at $T$ can be
understood qualitatively from Eq. (\ref{HSpart2}). For the present
case, $U_2<0$, the last term in Eq. (\ref{HSpart2}) favors the
alignment of the spins along the $x$-axis in the low $T$
phase. Consequently, the polarization, $S^z$, is always zero.

As in the $U_2 > 0$ case we observe a slow decay of the Green
functions $G_\sigma$ in the superfluid phase and the transition is
shown to be of the BKT type using the universal jump argument (see
Table \ref{tab2}).

\begin{table}
\begin{tabular}{ccc}
\hline
$t/U_0$ & $T_c \ (\rho_s)$ & $T_c \ (C_{\rm max}) $ \\
\hline
0.06 & $1.85 \pm 0.05$ & $1.75 \pm 0.10$ \\
0.07 & $2.00 \pm 0.05$ & $2.00  \pm 0.10$ \\
0.08 & $ 2.10 \pm 0.05 $ & $2.05 \pm 0.05$ \\
0.09 & $2.14 \pm 0.05$ & $ 2.16 \pm 0.05$\\
0.10 & $2.18 \pm 0.05$ & $ 2.15 \pm 0.05$\\
\hline
\end{tabular}
\caption{Values of the SF to NBL transition temperatures for different
  values of $t/U_0$ and for a $L=12$ size for $U_2/U_0=-0.1$ and
  $\rho=2$.  Similar to Table \ref{tab1}.
\label{tab2}}
\end{table}

\begin{figure}
\includegraphics[width=8.5cm]{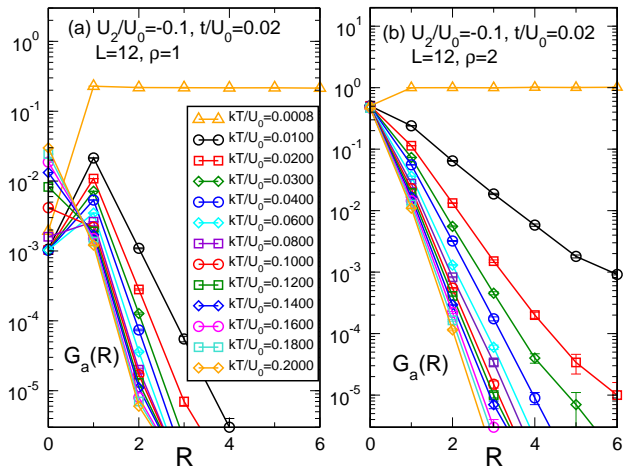}
\caption{(Color online) The anticorrelated Green functions in the
  $\rho=1$ (a) and $\rho=2$ (b) Mott regions for negative $U_2$. In
  both cases, the long range order of the anticorrelated movements
  that was present at $T=0$ is lost even for very small
  temperatures.\label{green_U2neg}}
\end{figure}

We also examined the anticorrelated Green function $G_{\rm a}$ in both
$\rho=1$ and $\rho=2$ MI regions at various temperatures. In the zero
temperature limit, both these phases exhibit nonzero values of $G_a$
at long distances. This is expected in both cases, considering the
pseudo-spin Hamiltonian. Neglecting the kinetic energy, for $U_2 < 0$,
the energy is minimized by maximizing $(S^x_{\bf r})^2$ on each
site. For $\rho=1$ and $\rho=2$, there are two degenerates states that
achieve this: the $S^x_{\bf r} = \pm 1/2$ for $\rho=1$ and the
$S^x_{\bf r} = \pm 1$ for $\rho=2$.  These degenerate ground states
are coupled by second order hopping processes which lift this
degeneracy through coherent anticorrelated movements
\cite{deforges11}. In Fig. \ref{green_U2neg}, we show $G_{\rm a}$ in
the $\rho=1$ and $\rho=2$ MI regions for different temperatures. As
expected, the phase coherence once again completely disappears rapidly
as the temperature is increased from zero and we do not observe any
sign of quasi long range phase coherence in the MI region, as well as
in the NBL phase.

\section{Conclusion}

Studying a bosonic spin-$1/2$ Hubbard model at finite temperature and
comparing to the $T=0$ case, we find that the effect of temperature is
dramatically different depending on the phase we consider.  The
superfluid phases are essentially unchanged by raising the
temperature: The long range order present at $T=0$ is transformed
into QLRO. On the other hand, the MI regions are drastically modified
as the polarization that occurs in certain cases is almost immediately
wiped out by thermal fluctuations: low temperature quasi polarized
states are not found in the regime of temperatures we studied.

This is due to the fact that the energy scales associated with the
polarization of the system take different values in these different
phases.  In the MI regions, the polarization is due to the coupling
between different low energy degenerate Mott states, as emphasized by
the pseudo-spin theory \cite{kuklov03}. These couplings are of order
$t^2/U_0$ and the pseudo-spin quasi-ordering vanishes very rapidly
with temperature.  On the other hand, in the superfluid, the energy
scale associated with the coupling of pseudo-spins is obviously much
larger. While it is not possible to specify this scale as precisely as
in the Mott phases, due to the itinerant nature of the particles in
the superfluid regime, a simple argument shows that it is of order
$U_2$: whenever the particles enter the superfluid phase and adopt
delocalized states, they overlap; there is an interaction cost which
is then of order $U_0$ for identical particles and $U_0 + U_2$ for
different ones. This favors having the particles mostly of the same
type for $U_2 > 0$ and leads to a quasi polarization. For $U_2 < 0$,
the interaction favors having a mixture of particles and gives a
system without any sign of polarization. In both cases, the energy
scale is typically $|U_2|$

The finite temperature phase diagram presented here is important for
the proper interpretation of experimental realization of this and
related systems using ultra-cold atoms loaded on optical
lattices. Such experiments are, of course, always at finite
temperature. Similarly to what happens for fermions, we observe that
the spin correlations in the Mott phases will be very difficult to
access experimentally as the associated energy scales are very small
and as the correlations are almost immediately wiped out by thermal
fluctuations. On the other hand, due to the relatively small sizes of
experimental systems and to its larger associated energies, the
quasi-ordering of spins in the superfluid phase should be immediately
visible experimentally and indistinguishable from true polarization of
the system.  In a finite size system, one would expect to observe a
slow drift of the polarization as the $yz$ symmetry of the system is
restored over time.

From a more technical point of view, we have elucidated the
limitations of the MFT commonly used in the literature. We found that
it is unable to distinguish correctly the MI and NBL regions. The MFT
does predict reasonably well the NBL-SF boundaries but not the nature
of the transition which is sometimes predicted to be of first order
whereas direct QMC simulations, and symmetry considerations, show that
it is in the BKT universality class.  Furthermore, we found that MFT
predicts direct first and second order transitions at finite $T$
between the MI and SF phases which QMC shows do not exist, (see
Figs. \ref{afrho1}, \ref{afrho2}, \ref{rho12_U2neg}).  Similar
incorrect MFT behavior was found for the spin-$1$ model and the same
caveats should be applied for example in Ref.\cite{Pai08}.

\begin{acknowledgments}
This work was supported by: the CNRS-UC Davis EPOCAL LIA joint
research grant; by NSF grant OISE-0952300; an ARO Award W911NF0710576
with funds from the DARPA OLE Program.
We would like to thank Michael Foss-Feig and Ana-Maria Rey for useful
input and discussion.
\end{acknowledgments}

\end{document}